# A RM-Polar codes

B. Li, H. Shen and D. Tse

In this letter we propose a new hybrid codes called "RM-Polar" codes. This new codes are constructed by combining the construction of Reed-Muller (RM) code and Polar code. It has much larger minimum Hamming distance than Polar codes, therefore it has much better error rate performance than Polar codes.

*Introduction:* Polar codes are a major breakthrough in coding theory [1]. They can achieve Shannon capacity with a simple encoder and a simple successive cancellation (SC) decoder, both with low complexity of the order of $O(N\log N)$, where $N$ is the code block size. But for short and moderate lengths, the error rate performance of polar codes with the SC decoding is not as good as LDPC or turbo codes. A new SC-list decoding algorithm was proposed for polar codes recently [2], which performs much better than the simple SC decoder and performs almost the same as the optimal ML (maximum likelihood) decoding at high SNR. In [3], we analyzed the minimum Hamming distance of the Polar codes, obtained a union bound on its frame error rate, and found that the performance of the Polar codes is dominated by its minimum Hamming distance. Although Polar codes can achieve the ML performance by the SC-List decoding, it still performs worse than turbo or LDPC codes due to its poor minimum Hamming distance. In this letter, we propose a hybrid codes called "RM-Polar" codes. This code is constructed by combining the code constructions of Reed-Muller codes and Polar codes. It not only has a larger minimum Hamming distance than the Polar codes, but also can be decoded as Polar codes by the SC decoder and the SC-List decoder. The simulations show that this RM-Polar code has much better error-rate performance than Polar codes.

*RM and Polar Codes:* Let $F = \begin{bmatrix} 1 & 0 \\ 1 & 1 \end{bmatrix}$, $F^{\otimes n}$ is a $N \times N$ matrix, where $N = 2^n$, $\otimes n$ denotes nth Kronecker power, and $F^{\otimes n} = F \otimes F^{\otimes (n-1)}$. Both Reed-Muller codes and Polar codes $(N, K)$, where K is the number of information bits, can be generated as

$$x = uF^{\otimes n} \quad (1)$$

where $x = (x_1, x_2, ..., x_N)$ is the encoded bit sequence, and $u = (u_1, u_2, ..., u_N)$ is the encoding bit sequence. The bit indexes of $u$ are divided into two subsets: the one containing the information bits represented by $A$ and the other containing the frozen bits represented by $A^c$. Since the frozen bits are usually set 0, then we have

$$x = u_A F^{\otimes n}(A) \quad (2)$$

where $F^{\otimes n}(A)$ denotes the submatrix of $F^{\otimes n}$ formed by the rows with indices in $A$, $u_A$ are the information bits, $K$ equals to the number of bits in $u_A$.

Although both RM codes and Polar codes are obtained from the same $F^{\otimes n}$ matrix, they select information bits according to different criteria. Polar codes select the information bits according to the Bhattacharyya parameters [1]. Firstly the Bhattacharyya parameters are computed for all of the bits $(u_1, u_2, ..., u_N)$, and then $K$ bits with the smallest Bhattacharyya parameters are selected as information bits, and the rest bits are selected as frozen bits. While the RM codes select the information bits according to the row weight. Firstly weights of all rows in $F^{\otimes n}$ matrix are computed, and then $K$ bits with the largest weights of their corresponding rows in $F^{\otimes n}$ matrix are selected as information bits, and the rest bits are selected as frozen bits.

*Minimum Hamming Distances of RM and Polar Codes:* For Polar codes, the information bits are selected to minimize error rate of the SC decoder. The minimum Hamming distance is not considered in the design of Polar codes, and it is shown that the minimum Hamming distance of the Polar code $(2048, 1024)$ is as low as 16 [3]. While RM codes have much larger minimum Hamming distance. The RM code $(N, K)$ has the following parameters: $K = \sum_{i=0}^{r} \binom{m}{i}$, $N = 2^m$, and its minimum Hamming distance is $d = 2^{m-r}$. If we choose $N = 2048$ and $r = 6$, we obtain a RM code $(2048, 1486)$. This RM code has a minimum Hamming distance $d = 32$ and the information bits of this RM code are with row-weights equal to or larger than 32.

Fig. 1 shows the row-weights of ordered bit indexes according to the Bhattacharyya parameters of the Polar code $(2048, 1024)$, where the first 1024 bits of the most-left part are with the smallest Bhattacharyya parameters and are selected as information bits. We can see that some of these selected information bits have the row-weights of 16, therefore the minimum Hamming distance of this Polar code is 16, which confirms the analysis in [3].

*Hybrid RM-Polar Codes:* We construct a new RM-Polar code $(2048, 1024)$ as follows. We first remove the bits with row-weights equal to or less than 16, and then select $K = 1024$ bits with the smallest Bhattacharyya parameters from the remaining bits. In other words, the information bits of this RM-Polar code are only selected from 1486 information bits of the RM code $(2048, 1486)$. Since the information bits of our newly designed RM-Polar code $(2048, 1024)$ are a subset of the information bits of the RM code $(2048, 1486)$, this RM-Polar code will have a minimum Hamming distance of 32.

Fig. 2 shows the frame error rate (FER) of the Polar code $(2048, 1024)$ and our new RM-Polar code $(2048, 1024)$ using the SC-List decoding, where the signal is BPSK modulated and transmitted over the additive white Gaussian noise (AWGN) channel. The ML bound is also simulated in the same way as in [2]. It is shown that the RM-Polar code performs much better than the Polar code and there is about 0.6dB gain of the RM-Polar code over the Polar code with the same list size $L = 32$. Although the SC-List decoding with $L = 32$ achieves the ML bound for the Polar code at high SNR, it is still far from the ML bound for the RM-Polar code. There is about 1.1dB difference between the ML bounds of the RM-Polar code and the Polar code at $FER = 10^{-3}$. When we increase the list size, we can continuously improve FER performance, and we can obtain around 0.4dB gain by using the list size $L = 2048$ over $L = 32$ at $FER = 10^{-3}$. The ML bound of this RM-Polar code achieves $FER = 10^{-3}$ at $E_b/N_o = 1.1 dB$, this is about 0.2dB from the information theoretic limit at this block length. But in order to achieve the ML bound, a very large list size is needed.

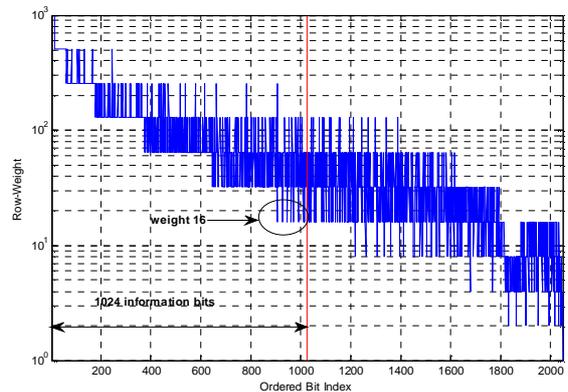

**Fig. 1** *The row-weights of ordered bit indexes according to Bhattacharyya parameters of the Polar code $(2048, 1024)$, where the most-left bit is with the smallest Bhattacharyya parameter.*

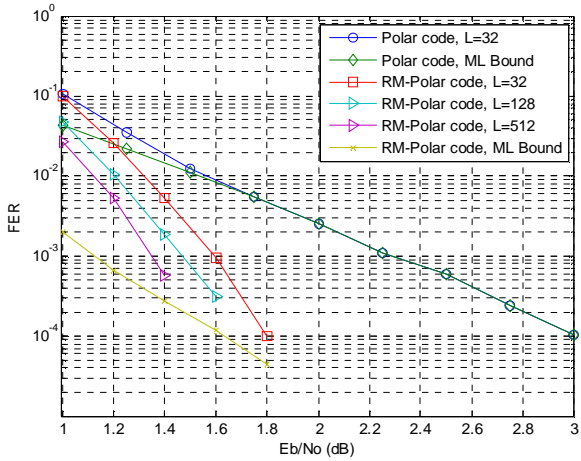

**Fig. 2** *Comparison of FER between the RM-Polar code* $(2048, 1024)$ *and the Polar code* $(2048, 1024)$ *under the SC-List decoder.*

*Conclusions:* In this letter, we propose a new "RM-Polar" codes. This code is constructed by using both the Bhattacharyya parameters used for Polar codes and row-weight used for RM code. The new code has much larger minimum Hamming distance than Polar codes, and can be decoded as Polar code using SC-List decoder. The simulations show that the new code has much better performance than Polar codes.

B. Li and H. Shen (Communications Technology Research Lab., Huawei Technologies, Shenzhen, P. R. China)
D. Tse (Dept. of Electrical Engineering and Computer Science, University of California at Berkeley, CA 94720-1170, USA ).
E-mail: binli.binli @huawei.com.